\documentclass[epj]{svjour}
\include{epsf}
\begin{document}
\title{Rheology of Semi-dilute Solutions of Calf-thymus DNA}
\author{Ranjini Bandyopadhyay and A. K. Sood}
\institute{Department of Physics, Indian Institute of Science, Bangalore 560 012, India}
\abstract{We study the rheology of semi-dilute solutions of the sodium salt of calf-thymus DNA in the linear and nonlinear regimes. The frequency response data can be fitted very well to the hybrid model with two dominant relaxation times $\tau_{\circ}$ and $\tau_{1}$. The ratio $\frac{\tau_{\circ}}{\tau_{1}} \sim$ 5 is seen to be fairly constant on changing the temperature from 20$^{\circ}$C to 30$^{\circ}$C. The shear rate dependence of viscosity can be fitted to the Carreau model.}
\titlerunning{Rheology of Semi-dilute Solutions of Calf-thymus DNA}
\authorrunning{R. Bandyopadhyay and A. K. Sood}
\date{}
\maketitle
\section{Introduction}

Deoxyribonucleic acid (DNA) is a key constituent of the nucleus of living cells and is composed of building blocks called nucleotides consisting of deoxyribose sugar, a phosphate group and  four nitrogenous bases - adenine, thymine, guanine and cytosine \cite{stryer}. X-ray crystallography shows that a DNA molecule is shaped like a double helix, very much like a twisted ladder \cite{watson}. The ability of DNA to contain and transmit genetic information makes it a very important biopolymer that has been the subject of intense scientific research in recent years.  DNA macromolecules are charged and depending on their molecular weight, the conformation could vary between a rigid rod and a flexible coil. The ability of flexible DNA molecules to twist, bend and change their conformation under tension or shear flow have been extensively studied both theoretically and experimentally \cite{marko,leduc}. The linear viscoelastic moduli of calf-thymus DNA solutions have been measured by Mason {\it et al.} at room temperature in the concentration range 1-10 mg/ml \cite{mason}. They have noted that the measured viscoelastic spectra do not fit the standard reptation model for flexible polymers \cite{degennes}. Measurements of the nonlinear rheology of entangled T4 Bacteriophage DNA molecules \cite{jary} show a plateau region in the stress $\sigma$ after an initial Newtonian regime at very low shear rates $\dot\gamma$, similar to the flow curves seen in surfactant gels \cite{spenley,ranpap}. 

\noindent The highly charged DNA macromolecule is a typical  polyelectrolyte and the effects of the long-range and intra-chain Coulomb interactions on its structure and dynamics are significant. The structure of short DNA fragments in aqueous salt solutions have been studied using small-angle neutron scattering \cite{maarel}. Direct mechanical measurements of the elasticity of single $\lambda$-DNA molecules show deviations from the force curves predicted by the freely-jointed  chain model \cite{bueche}, leading to the conclusion that DNA in solution has significant local curvature \cite{smith}. Measurements of relaxation times of a single DNA molecule manipulated using laser tweezers and observed by optical microscopy \cite{perkins} show a qualitative agreement with the dynamic scaling predictions of the Zimm model \cite{zimm}. In recent years, experiments have been carried out on the electrohydrodynamic instability observed in DNA solutions under the action of a strong electric field \cite{isambert}. Electric fields of the proper frequency and amplitude lead to the formation of islands of circulating molecules, with the islands arranging into a herring-bone formation. Recent simulation studies on short, supercoiled DNA chains show that DNA is a glassy system with numerous local energy minima under suitable conditions \cite{zhang}. 

\noindent  \noindent In this paper, we discuss our recent results on the linear and nonlinear rheology of semi-dilute solutions of the sodium salt of calf-thymus DNA. We fit our frequency response data to the hybrid model \cite{warren} and the shear viscosity {\it vs.} shear rate data to the Carreau model \cite{carreau}. These models are essentially for dilute polymer solutions and the value of the radius of gyration $R_{g}$ of the DNA macromolecules calculated from the fits to the hybrid model \cite{warren} are highly overestimated. We believe that the polydispersity of the DNA chains, the overlap of the macromolecules in the semi-dilute concentration regime and the electrostatic interaction between the ionized groups also need to be considered for a more accurate theoretical description of our experimental results. 

\section{The hybrid model theory}

The hybrid model theory incorporating Zimm dynamics has been used in the dilute polyelectrolyte and polymer literature to explain the behaviour of the reduced elastic modulus $G_{R}^{\prime}(\omega)$ =$\lim_{c \rightarrow 0}G^{\prime}(\omega)$ and the reduced viscous modulus $G_{R}^{\prime\prime}(\omega)$ =$\lim_{c \rightarrow 0}[G^{\prime\prime}(\omega)-\omega\eta_{s}$] observed in aqueous solutions of sodium poly(styrene sulphonate) \cite{rosser} and dilute aqueous samples of polymers such as separan AP-30, xanthan gum, carboxymethyl cellulose and Polyox WSR301 \cite{tam}. Here, $\eta_{s}$ is the solvent viscosity and $c$ is the polymer concentration. The hybrid model is characterised by a series of relaxation times spaced according to the Zimm theory \cite{zimm}, together with one additional longer relaxation time $\tau_{\circ}$. The Zimm model treats a single polymer chain in the framework of a bead-spring model, with N beads connected by N+1 springs, in the presence of hydrodynamic interactions. The hybrid model is characterised by a series of relaxation times spaced as in the Zimm theory \cite{zimm}, in addition to a longest relaxation time $\tau_{\circ}$. In the limit of infinite dilution, the intrinsic moduli for this model may be written as \cite{warren} \\
$G^{\prime}_{R}(\omega)=G_{\circ}\omega^{2}\tau_{\circ}^2[1+\omega^{2}\tau_{\circ}^2]^{-1}+ G_{1}\Sigma_{p=1}^{N}\frac{\omega^{2}\tau_{1}^2(\frac{\tau_{p}^{2}}{\tau_{1}^{2}})}{1+\omega^{2}\tau_{1}^2(\frac{\tau_{p}^{2}}{\tau_{1}^{2}})}$ 
 [1]\\
$G^{\prime\prime}_{R}(\omega)=  \omega\tau_{\circ}[G_{\circ}(1+\omega^{2}\tau_{\circ}^2)^{-1}+m_{2}] + G_{1}\Sigma_{p=1}^{N}\frac{\omega\tau_{1}(\frac{\tau_{p}}{\tau_{1}})}{1+\omega^{2}\tau_{1}^2(\frac{\tau_{p}^{2}}{\tau_{1}^{2}})}$ \\ 
\hspace{17cm} [2],\\
where
$\tau_{p}=\frac{\tau_{1}}{p^{3\nu}}$, $p$ = 1, 3, 5 ... and $3\nu$=1.66 according to the Zimm model. We recall that in the Zimm model, only the time scales $\tau_{p}$, which correspond to internal motions such as flexure of the chain, contribute to the stress relaxation process.  The experimental data for the elastic and viscous moduli of poly(2-vinyl pyridene) \cite{hodgson} have been fitted to the predictions of the Zimm model. The peaks in the viscoelastic moduli observed as a function of polyelectrolyte concentration have been explained in terms of electrostatically driven polymer coil expansion and contraction. Computer simulations have shown that on increasing the charge on the macromolecule, the longest relaxation time $\tau_{1}$ in the Zimm spectrum \cite{wada3} is enhanced.

\noindent The hybrid model was proposed by Warren {\it et al.} to explain the infinite-dilution viscoelastic properties of the helical molecule PBLG (poly-$\gamma$- benzyl -L -glutamate) \cite{warren}. The value of $\tau_{\circ}$ for infinitely dilute solutions of PBLG, obtained from the fits of $G_{R}^{\prime}$ and $G_{R}^{\prime\prime}$ to the Eqns. 1 and 2 respectively, has been associated with the rotational diffusion of the macromolecule. The fitted value of $\tau_{1}$ has been explained in terms of the flexural modes of vibration of the helix damped by solvent viscosity. Okamoto {\it et al.} have explained the viscoelastic properties of dilute solutions of the polyelectrolytes poly (acrylic acid) and poly (methacrylic acid) using this model, but without extrapolation to infinite dilution \cite{wada2,wada1}. The authors conclude that the relaxation of these polyelectrolytes involves the rotation of the whole macromolecule, together with internal configurational changes.
\section{Experiment} 

We have carried out rheometric measurements on semi-dilute solutions of the sodium salt of calf-thymus DNA, dissolved in Tris-EDTA buffer (pH adjusted to physiological conditions of 7.9) at a concentration of 1mg/ml (overlap concentration of calf-thymus DNA is c$^{\star}$=0.35 mg/ml for a good solvent). The DNA was purchased from Sigma Chemicals, India in lyophilized form. Calf-thymus DNA is a linear, double-stranded macromolecule, and consists of $\sim$ 13,000 base pairs. Since 1 nucleotide $\equiv$ 324.5 Da, the estimated molecular weight of calf-thymus DNA is $\sim$ 8.4$\times$10$^{7}$ Da. We have dispersed the lyophilized DNA powder in an aqueous buffer consisting of  10mM Tris,  100mM NaCl, 50mM each of NaCl and KCl and 5 mM MgCl$_{2}$. The samples prepared in this way were allowed to equilibrate for a day at 4$^{\circ}$C to inhibit degradation. The rheological measurements have been conducted at 20$^{\circ}$C, 25$^{\circ}$C and 30$^{\circ}$C in an AR-1000N Rheolyst stress controlled rheometer (T. A. Instruments, U. K.) equipped with temperature control and software for shear rate control. We have used a stainless steel cone-and-plate assembly with a diameter of 4 cm and an angle of 1$^{\circ}$59' as the shearing geometry. All the experiments reported below have been carried out on a sample of concentration 1 mg/ml. Because all experiments have been performed on the same sample, we have allowed sufficient time between runs to ensure that the sample relaxes fully to its equilibrium state before the start of the subsequent experiment. For the oscillatory experiments, the rheometer has the provision for showing the waveform of the response on the application of an oscillatory stress. Care has been taken to ensure a distortion-free response to the applied oscillatory stresses for all the experiments.
 
\section{Results}
In this section, we discuss our experimental results on the linear and nonlinear rheology of calf-thymus DNA (concentration $c$=1mg/ml) at 20$^{\circ}$C, 25$^{\circ}$C and 30$^{\circ}$C.

\subsection{Linear rheology} 

Figs. 1, 2 and 3 show the frequency response data ({\it i.e.} the measured $G^{\prime}(\omega)$ and [$G^{\prime\prime}(\omega)-\omega\eta_{s}$] {\it vs.} the angular frequency $\omega$, where the solvent viscosity $\eta_{s}$ = 1$\times$10$^{-3}$ Pa) at 20$^{\circ}$C, 25$^{\circ}$C and 30$^{\circ}$C and the corresponding fits (shown by solid lines) to the hybrid model with $p$ =1 (Eqns. 1 and 2) \cite{warren} over almost three decades of angular frequency. Insets of figs. 1, 2 and 3 show the calculation of the oscillatory stress $\sigma_{osc}$ lying in the linear regime. Because $G^{\prime}(\omega)$ and $G^{\prime\prime}(\omega)$ do not change appreciably over the range 8mPa to 30mPa at an oscillatory frequency $\omega$ of 0.628 rads$^{-1}$, we have controlled the oscillatory stress for subsequent frequency response experiments at $\sigma_{osc}$ = 25mPa.
\begin{figure}
\centerline{\epsfxsize = 9cm \epsfysize = 7cm \epsfbox{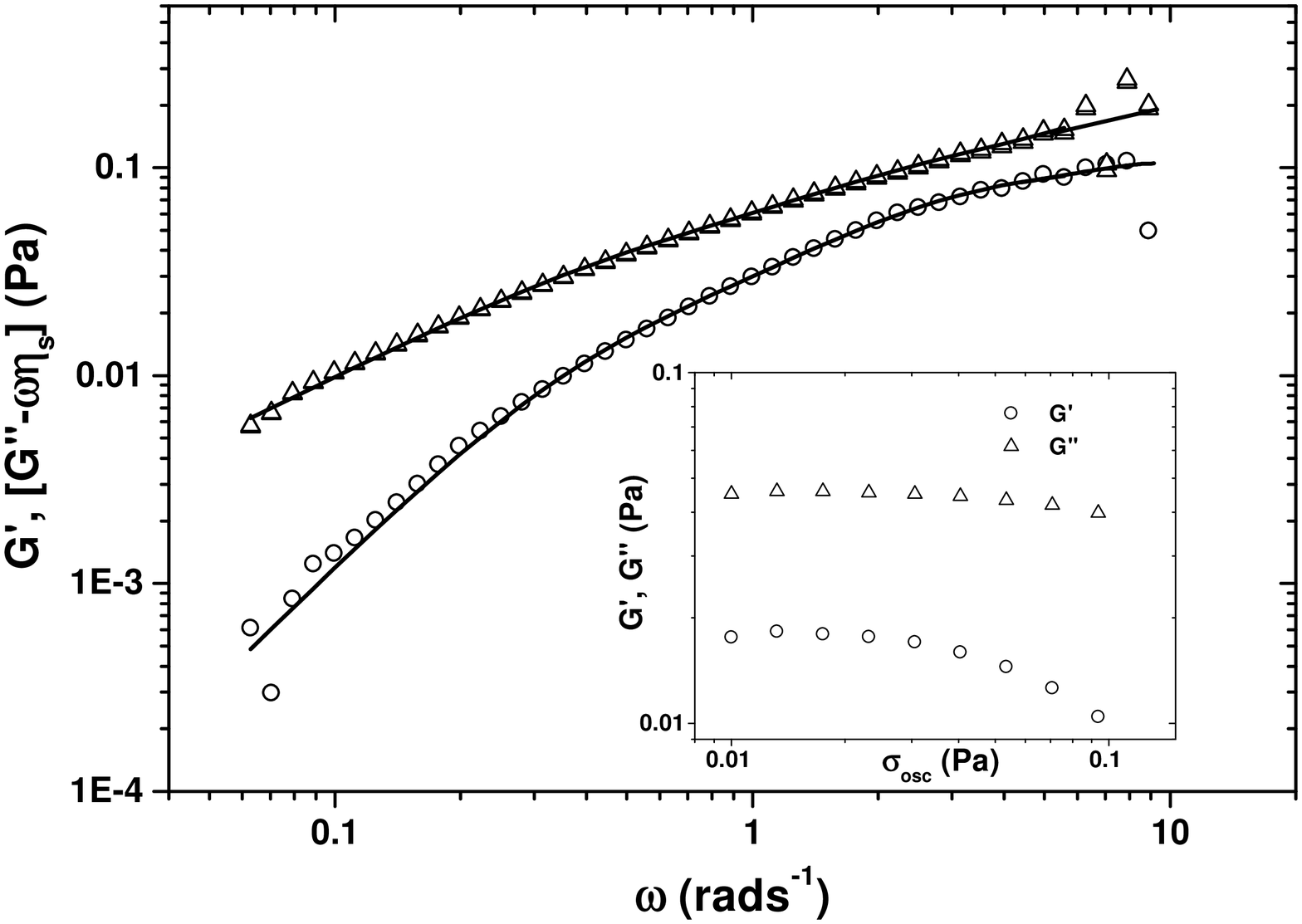}}
\caption{Elastic modulus $G^{\prime}(\omega)$ (open circles) and viscous modulus [$G^{\prime\prime}(\omega)-\omega\eta_{s}$] (open triangles) {\it vs.} angular frequency $\omega$ at T=20$^{\circ}$C and $\sigma_{osc}$ = 0.025Pa. The solid lines show the fits to the hybrid model \cite{warren}. The inset shows the plot of $G^{\prime}(\omega)$ and $G^{\prime\prime}(\omega)$ {\it vs.} the oscillatory stress $\sigma_{osc}$ at $\omega$ = 0.628 rads$^{-1}$.}
\end{figure}
\begin{figure}
\centerline{\epsfxsize = 9cm \epsfysize = 7cm \epsfbox{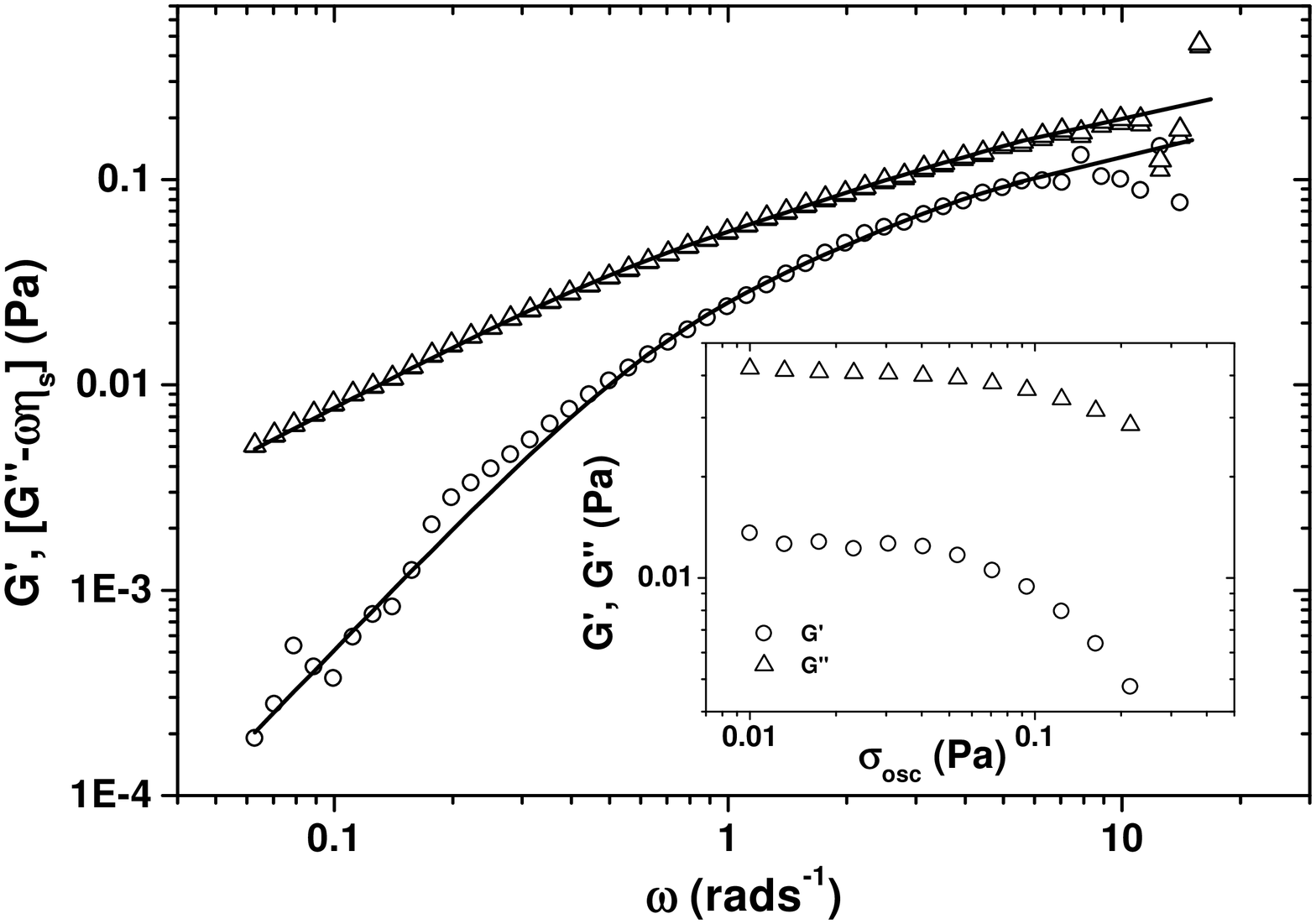}}
\caption{Elastic modulus $G^{\prime}(\omega)$ (open circles) and the viscous modulus [$G^{\prime\prime}(\omega)-\omega\eta_{s}$] (open triangles) {\it vs.} angular frequency $\omega$ at T=25$^{\circ}$C and $\sigma_{osc}$ = 0.025Pa. The solid lines show the fits to the hybrid model \cite{warren}. The inset shows the plot of $G^{\prime}(\omega)$ and $G^{\prime\prime}(\omega)$ {\it vs.} the oscillatory stress $\sigma_{osc}$ at $\omega$ = 0.628 rads$^{-1}$.}
\end{figure}
\begin{figure}
\centerline{\epsfxsize = 9cm \epsfysize = 7cm \epsfbox{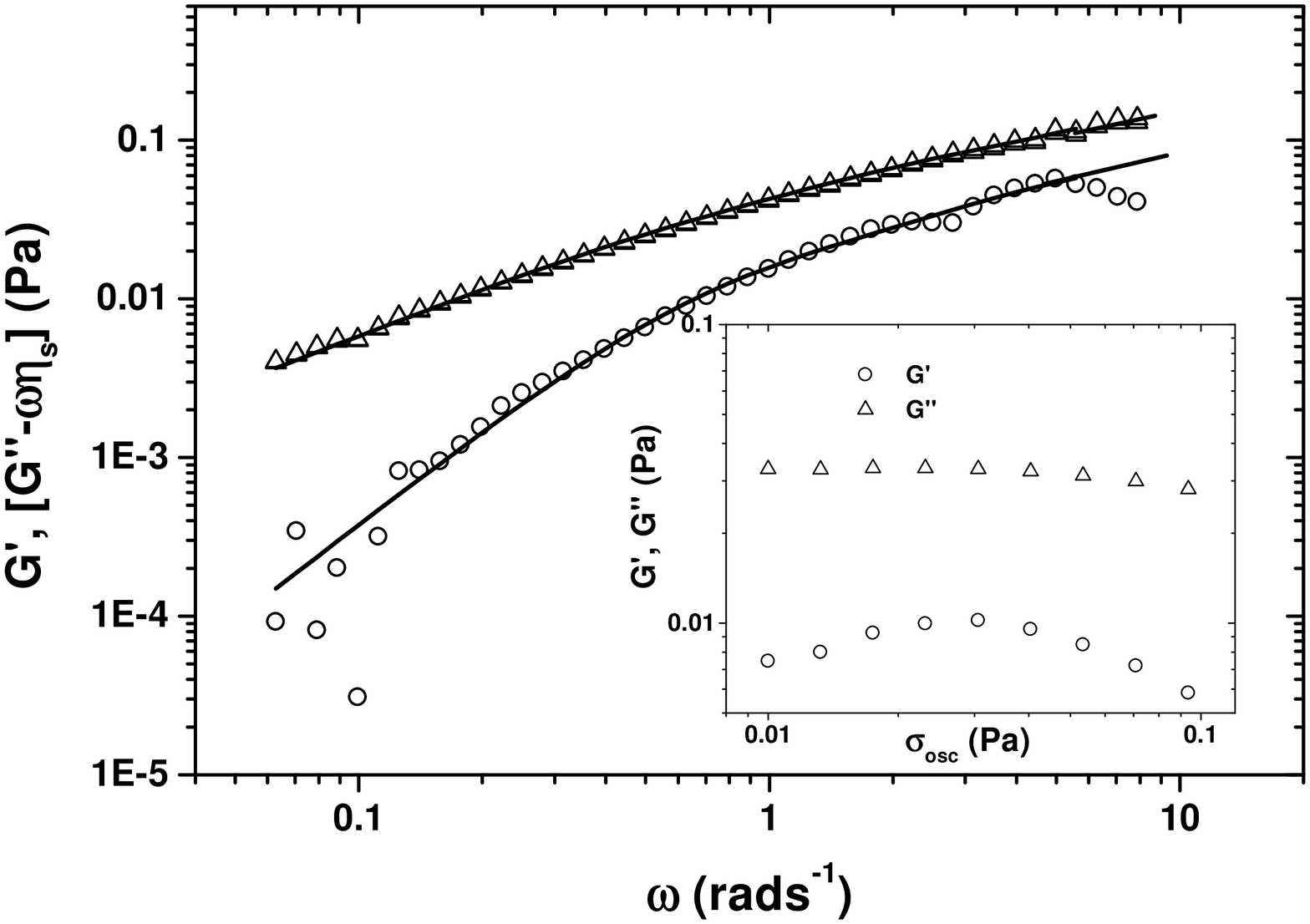}}
\caption{Elastic modulus $G^{\prime}(\omega)$ (open circles) and viscous modulus [$G^{\prime\prime}(\omega)-\omega\eta_{s}$] (open triangles) {\it vs.} angular frequency $\omega$ at T=30$^{\circ}$C and $\sigma_{osc}$ = 0.025Pa. The solid lines show the fits to the hybrid model \cite{warren}. The inset shows the plot of $G^{\prime}(\omega)$ and $G^{\prime\prime}(\omega)$ {\it vs.} the oscillatory stress $\sigma_{osc}$ at $\omega$ = 0.628 rads$^{-1}$.}
\end{figure}
The values of the fitting parameters $G_{\circ}$, $\tau_{\circ}$, $G_{1}$, $\tau_{1}$ and $m_{2}$ are shown in Table 1. The measured $G^{\prime}(\omega)$ and [$G^{\prime\prime}(\omega)-\omega\eta_{s}$] could not be fitted to the Tanaka model \cite{tanaka} for rigid rods given by $G_{R}^{\prime}(\omega) = G_{\circ}\omega^{2}\tau_{\circ}^2(1+\omega^2\tau_{\circ}^{2})^{-1}$ and $G_{R}^{\prime\prime}(\omega) = \omega\tau_{\circ}[G_{\circ}(1+\omega^2\tau_{\circ}^{2})^{-1}+m_{2}]$ over the entire frequency range for the frequency response data corresponding to T=20$^{\circ}$C, 25$^{\circ}$C and 30$^{\circ}$C \cite{phdran}.
It has been pointed out earlier by Mason {\it et al.} \cite{mason} that the frequency response data for calf-thymus DNA could not be fitted to the standard models of reptation dynamics. We also find that the data for $G^{\prime}(\omega)$ and $G^{\prime\prime}(\omega)$  \cite{phdran} cannot be fitted to the Doi-Edwards model \cite{degennes,doi} for flexible polymers. 
\begin{table}
\begin{center}
\caption{Values of the fitting parameters $G_{\circ}$, $\tau_{\circ}$, $G_{1}$, $\tau_{1}$ and $m_{2}$ obtained by fitting the hybrid model to $G^{\prime}(\omega)$ and [$G^{\prime\prime}(\omega)-\omega\eta_{s}$] respectively at T=20$^{\circ}$C, 25$^{\circ}$C and 30$^{\circ}$C.}
\vspace{0.5cm}
\begin{tabular}{|c|c|c|c|c|c|c|}\hline
T $^{\circ}$C & &$\tau_{\circ} (s)$ &$G_{\circ} (Pa)$  &$\tau_{1} (s)$ &$G_{1} (Pa)$ &$m_{2} (Pa)$\\ \hline
20  &$G^{\prime}$ &2.33 &0.021 &0.430 &0.080 & \\ \hline
20  &$G^{\prime\prime}$ &1.75 &0.027 &0.330 &0.096 &0.326  \\ \hline\hline
25  &$G^{\prime}$ &1.15 &0.0344 &0.246 &0.097 & \\ \hline
25  &$G^{\prime\prime}$ &1.140 &0.0386 &0.234 &0.1028 &0.383 \\ \hline\hline
30  &$G^{\prime}$ &1.31 &0.021 &0.230 &0.062 & \\ \hline
30  &$G^{\prime\prime}$ &1.14 &0.020 &0.22 &0.069 &0.335 \\ \hline\hline
\end{tabular}
\end{center}
\end{table}
 
\begin{figure}
\centerline{\epsfxsize = 10cm \epsfysize = 8cm \epsfbox{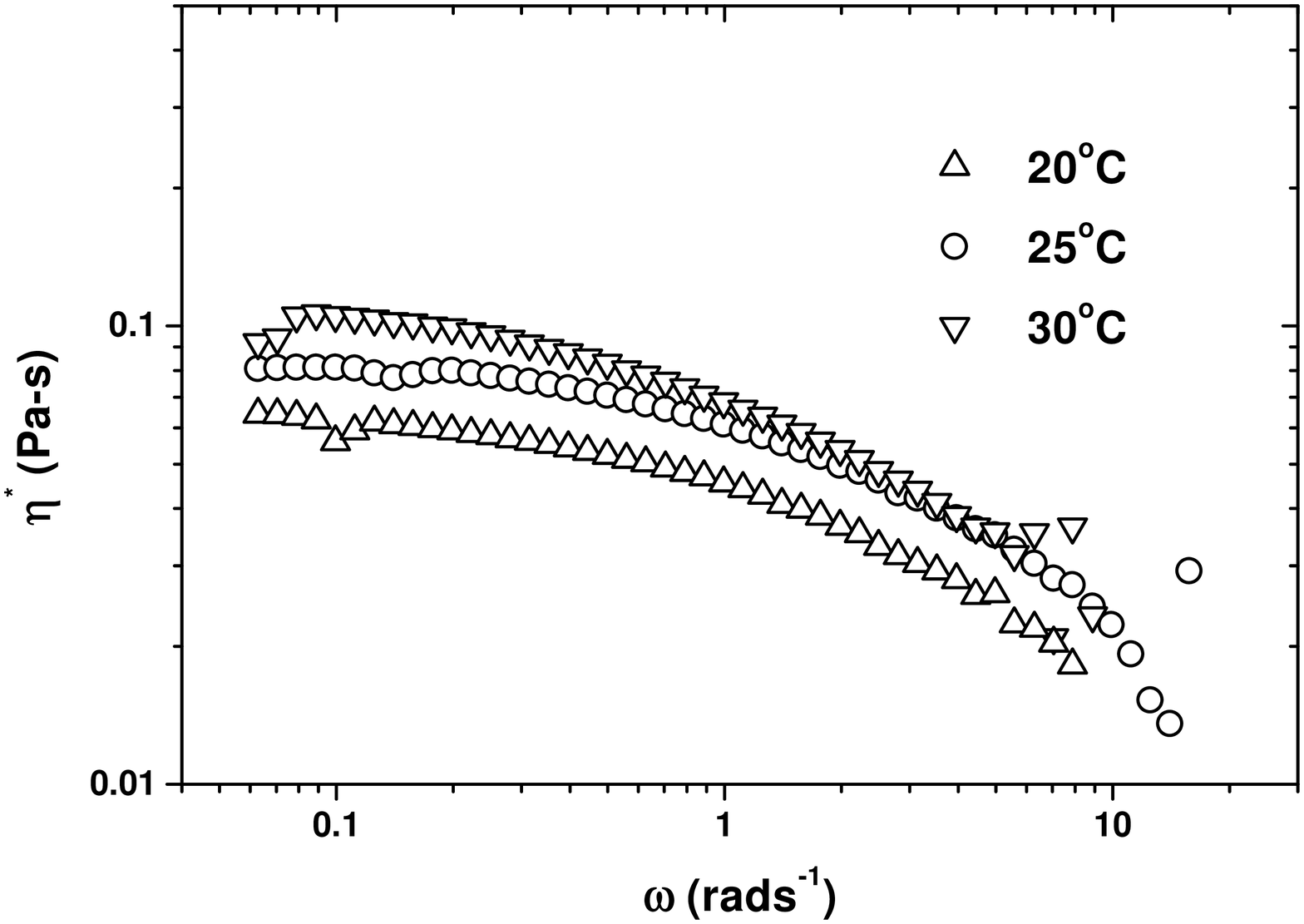}}
\caption{Dynamic viscosity $\eta^{\star}(\omega)$ {\it vs.} angular frequency $\omega$ at T=20$^{\circ}$C (up-triangles), 25$^{\circ}$C (circles) and 30$^{\circ}$C (down-triangles).} 
\end{figure}
In fig. 4, we have plotted the dynamic viscosity $\eta^{\star}(\omega)$ {\it vs.} ${\omega}$, where $\eta^{\star}(\omega) = \frac{(G^{\prime 2} + G^{\prime\prime 2})^{0.5}}{\omega}$ at 20$^{\circ}$C (up-triangles), 25$^{\circ}$C (circles) and 30$^{\circ}$C (down-triangles). 

\subsection{Nonlinear rheology}
\begin{table}
\begin{center}
\caption{Values of the fitting parameters $\eta_{\circ}$, $\tau_{R}$ and $m$ obtained by fitting the Carreau model \cite{carreau} to $\eta(\dot\gamma)$ {\it vs.} $\dot\gamma$  at T=20$^{\circ}$C, 25$^{\circ}$C and 30$^{\circ}$C, respectively.}
\vspace{0.5cm}
\begin{tabular}{|c|c|c|c|}\hline
T $^{\circ}$C &$\eta_{\circ}$ (Pa-s) &$\tau_{R}$ (s) & $m$\\ \hline \hline
{\rule[-3mm]{0mm}{8mm} 20}  &0.060$\pm$0.002 &0.499$\pm$0.145 &0.2425 \\ \hline \hline
{\rule[-3mm]{0mm}{8mm} 25}  &0.053$\pm$0.002 &0.34$\pm$0.11 &0.2422 \\ \hline \hline
{\rule[-3mm]{0mm}{8mm} 30}  &0.041$\pm$0.08 &0.36$\pm$0.08 &0.2430 \\ \hline \hline
\end{tabular}
\end{center}
\end{table}
\begin{figure}
\centerline{\epsfxsize = 9cm \epsfysize = 7cm \epsfbox{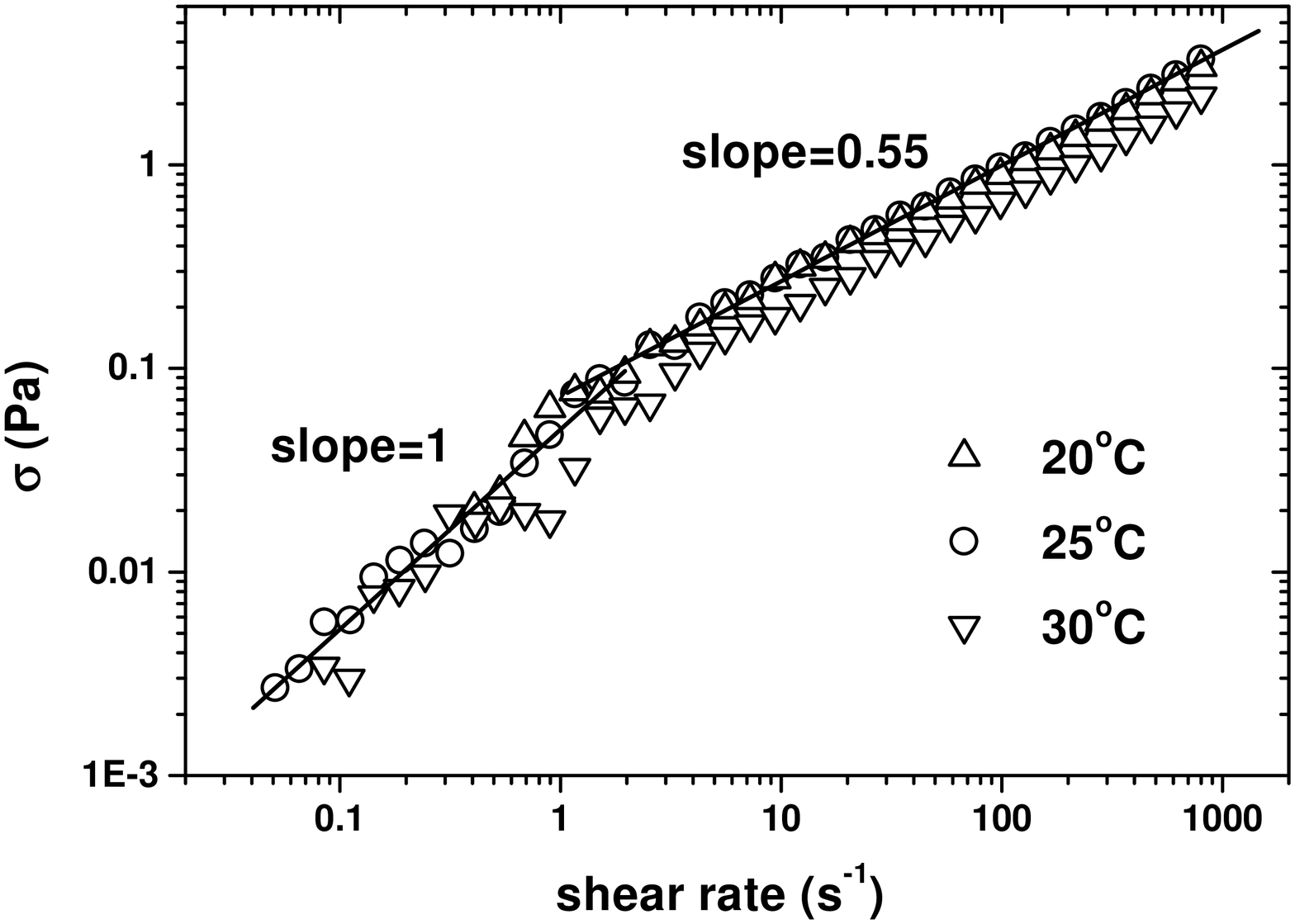}}
\caption{Plot of the stress $\sigma$ {\it vs.} shear rate $\dot\gamma$ at T=20$^{\circ}$C (denoted by up-triangles), 25$^{\circ}$C (circles) and 30$^{\circ}$C (down triangles). The solid lines show the linear fits to the plots in the two regimes given by $\dot\gamma <$ 2 and $\dot\gamma \ge$2. Note that at $\dot\gamma <$ 2, the slope of the fit is very close to 1.}
\end{figure}
\begin{figure}
\centerline{\epsfxsize = 9cm \epsfysize = 12cm \epsfbox{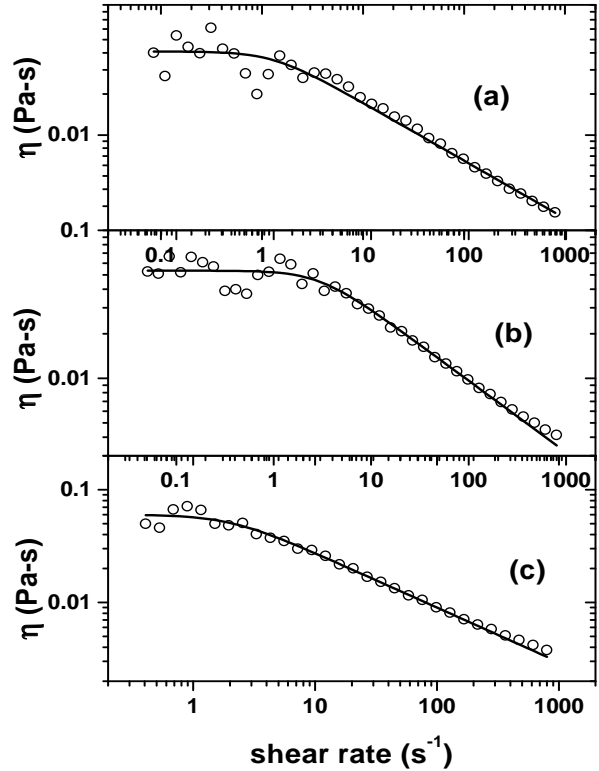}}
\caption{Plot of the shear viscosity $\eta(\dot\gamma)$ {\it vs.} shear rate $\dot\gamma$ at (a) T=20$^{\circ}$C, (b) 25$^{\circ}$C and (c) 30$^{\circ}$C. The solid lines are the fits to the Carreau model.}
\end{figure}
Fig 5 shows the plot of stress $\sigma$ versus shear rate $\dot\gamma$ at 20$^{\circ}$C (up triangles), 25$^{\circ}$C (circles) and 30$^{\circ}$C (down triangles) respectively. Initially, upto $\dot\gamma \sim$ 2s$^{-1}$, the slope of the flow curve is  very close to the Newtonian value of 1. At $\dot\gamma \ge$ 2$s^{-1}$, shear-thinning occurs and the fits to $\sigma \sim \dot\gamma^{\alpha}$ give $\alpha$ = 0.55. 
In fig. 6, we have plotted the shear viscosity $\eta(\dot\gamma)$ versus shear rate $\dot\gamma$ at (a) 20$^{\circ}$C, (b) 25$^{\circ}$C and (c) 30$^{\circ}$C. The plots of viscosity are found to fit well to the Carreau model \cite{carreau} written as $\eta(\dot\gamma) = \frac{\eta_{\circ}}{(1 + {\dot\gamma}^{2}\tau_{R}^2)^{m}}$. The values of the parameters obtained by fitting the shear viscosity data to the Carreau model have been listed in Table 2. Note that the values of $\eta_{\circ}$ obtained from these fits are found to decrease with increasing temperature. However, the temperature window is not wide enough to ascertain a possible Arrhenius dependence of the viscosities.

\section{Discussions of the results} 
\begin{figure}
\centerline{\epsfxsize = 9cm \epsfysize = 12cm \epsfbox{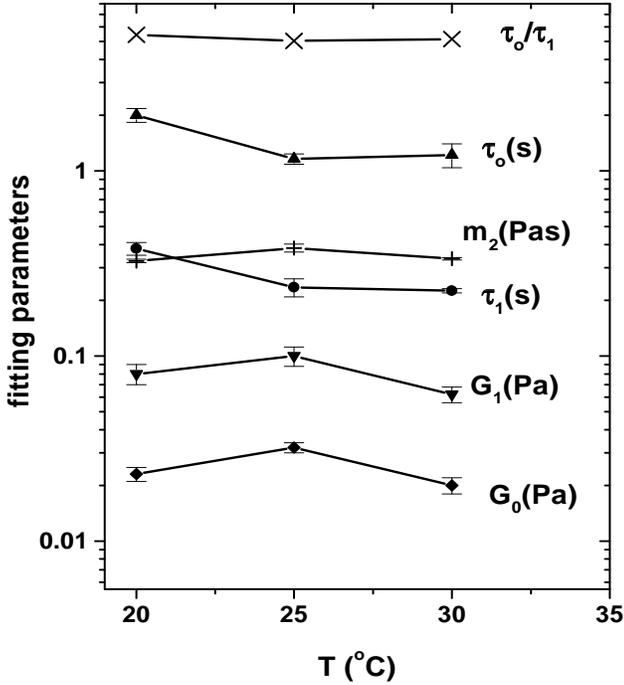}}
\caption{Plot of the fitting parameters $G_{\circ}$, $G_{1}$, $\tau_{\circ}$, $\tau_{1}$ and $m_{2}$ obtained from the fits to the hybrid model \cite{warren} of the frequency response data at T=20$^{\circ}$C, 25$^{\circ}$C and 30$^{\circ}$C. The ratio $\frac{\tau_{\circ}}{\tau_{1}}$, which is also plotted, shows a negligible variation with temperature. }
\end{figure}

Fig. 7 shows the plots of the fitting parameters $G_{\circ}$, $G_{1}$,
$\tau_{\circ}$, $\tau_{1}$ and $m_{2}$ obtained by fitting the frequency
response data at T=20$^{\circ}$C, 25$^{\circ}$C and 30$^{\circ}$C to the hybrid
model (Eqns. 1 and 2) {\it vs.} the temperature. It is seen that the relaxation
times $\tau_{\circ}$ and $\tau_{1}$ decrease monotonically on increasing
temperature, while the percentage changes in $G_{\circ}$ and $G_{1}$ are less
significant than the changes in $\tau_{\circ}$ and $\tau_{1}$ (Table
1). $\tau_{\circ}$ may be associated with the rotational diffusion of each DNA
macromolecule as a whole, whereas the time scale $\tau_{1}$, which corresponds
to the internal degrees of freedom of DNA, might signify the time scale of the
longest flexural mode of the macromolecule. The ratio of the two relaxation
times $\frac{\tau_{\circ}}{\tau_{1}} \sim 5$ is found to be fairly independent
of the temperature. In an experimental study on the helical macromolecule PBLG
\cite{warren}, it has been shown that the relaxation spectrum, which may be
fitted to the hybrid model, consists of a dominant relaxation time
$\tau_{\circ}$ describing the end-over-end rotation of the macromolecule, and a
time $\tau_{1}$ associated with its flexural bending mode. Assuming calf-thymus
DNA to be a flexible coil undergoing predominantly rotational diffusion, its
radius of gyration estimated from the relation
$\tau_{\circ}=0.325\times6^{\frac{3}{2}}\frac{\eta_{s}R_{g}}{k_{B}T}$
\cite{doi} is found to be 1.6$\mu$m at 25$^{\circ}$C. This value overestimates
the predicted value of $R_{g} \sim$ 0.4$\mu$m for calf-thymus DNA \cite{mason}.
For a rigid polymer, the rotational relaxation time $\tau_{\circ}$ may be
written as $\tau_{\circ}=\frac{\pi\eta_{s}L^{3}}{18k_{B}T}[\ln(\frac{2L}{d})]$
\cite{ulman}, where $L$ is the length of the polymer and $d$ is its diameter.
Taking $\frac{2L}{d}$=100, a rotational relaxation time of 1 s at 25$^{\circ}$C
gives an estimate of the length $L \sim$ 2$\mu$m. Owing to the large
configurational entropy of DNA and the fact that the DNA is in the semi-dilute
concentration regime, the possibility of DNA macromolecules existing as rigid
rods is extremely remote. This is further confirmed by the poor fit of the
frequency response data to the Tanaka model for rigid rods \cite{tanaka}. The
reason behind the discrepancy of the calculated $R_{g}$ with other estimates
\cite{mason} can be due to the application of the hybrid model to a finite
concentration of DNA, this model being ideally applicable only in cases of
infinite dilution. Clearly, the effects of the overlap of the DNA
macromolecules need to be incorporated in the theory. The Rouse model
\cite{doi} rather than the Zimm model might be a more suitable model to
describe the results. We recall that in the Rouse model, the time-scales in the
relaxation spectrum may be expressed as $\tau_{p} \sim \frac{\tau_{1}}{p^{2}}$,
where $p$ =1, 3, 5 etc. The polydispersity of the DNA chains and the
electrostatic interaction between the ionized surface groups also need to be
considered for an accurate description of our results. In spite of these
shortcomings in our explanation of $\tau_{\circ}$ and $\tau_{1}$, we note that
the frequency response data fits remarkably well to the hybrid model over
almost three decades of angular frequency.

\noindent Dilute solutions of rod-like macromolecules are known to show a power-law shear-thinning after an initial Newtonian regime at low $\dot\gamma$ \cite{tam,chaffey}. The viscosity data can be fitted to the Carreau's model \cite{carreau}, with the fitted values of $\eta_{\circ}$ increasing with decrease in temperature. Power-law shear-thinning in nonlinear flow experiments and Zimm dynamics in the linear rheology data has been previously observed in dilute solutions of polymers such as Separan AP30, xanthan gum {\it etc.} by Tam {\it et al.} \cite{tam}.

\section{Conclusions}
We have studied the linear and nonlinear rheology of semi-dilute solutions of calf-thymus DNA. The frequency response data show excellent fits to the hybrid model proposed by Warren {\it et al.} \cite{warren}. In order to understand clearly the physical significance of $\tau_{\circ}$ and $\tau_{1}$, we need to extend the results of the hybrid model to the case of semi-dilute polylectrolytes. The electrostatic repulsion between ionisable groups and the polydispersity of DNA macromolecules need to be accounted for. The flow curves at all temperatures may be divided into two regimes: an initial Newtonian regime at $\dot\gamma <$ 2s$^{-1}$ followed by a region of shear thinning at $\dot\gamma \ge$ 2s$^{-1}$, as indicated by $\alpha$ = 0.55 in the fits to $\sigma \sim \dot\gamma^{0.55}$ at all temperatures. The shear viscosities $\eta(\dot\gamma$) measured in these experiments and plotted {\it vs.} $\dot\gamma$ can be fitted to the Carreau model \cite{carreau}. It will be interesting to study the rheology of this system as a function of DNA concentration, in order to understand the changes in the rheological parameters across the overlap concentration $c^{\star}$.

\end{document}